\documentstyle[psbox]{WORKSHOP}
\begin{document}

\newcommand{\chandra}{{\it Chandra}}
\newcommand{\asca}{{\it ASCA}}
\newcommand{\rosat}{{\it ROSAT}}
\newcommand{\sax}{{\it BeppoSAX}}
\newcommand{\xmm}{{\it XMM-Newton}}
\newcommand{\swift}{{\it Swift}}
\newcommand{\ho}{Holmberg IX X-1}
\newcommand{\ngc}{NGC\,5408 X-1}
\newcommand{\eso}{ESO\,243-49}
\newcommand{\lum}{\thinspace\hbox{$\hbox{erg}\thinspace\hbox{s}^{-1}$}}
\newcommand{\flux}{\thinspace\hbox{$\hbox{erg}\thinspace\hbox{cm}^{-2}\thinspace\hbox{s}^{-1}$}}

\def\spose#1{\hbox to 0pt{#1\hss}}
\def\laeq{\mathrel{\spose{\lower 3pt\hbox{$\mathchar"218$}}
     \raise 2.0pt\hbox{$\mathchar"13C$}}}
\def\gaeq{\mathrel{\spose{\lower 3pt\hbox{$\mathchar"218$}}
     \raise 2.0pt\hbox{$\mathchar"13E$}}}

\title{
Monitoring Variable X-ray Sources in Nearby Galaxies \\ 
}

\author{
Albert~K.~H.~Kong
\\[12pt]  
%
Institute of Astronomy, National Tsing Hua University, Hsinchu, Taiwan \\
%
{\it E-mail: akong@phys.nthu.edu.tw} 
}

\abst{In the last decade, it has been possible to monitor variable X-ray sources in nearby galaxies. In particular, since the launch of {\it Chandra}, M31 has been regularly observed. It is perhaps the only nearby galaxy which is observed by an X-ray telescope regularly throughout operation. With 10 years of observations, the center of M31 has been observed with {\it Chandra} for over 1 Msec and the X-ray skies of M31 consist of many transients and variables. Furthermore, the X-ray Telescope of {\it Swift} has been monitoring several ultraluminous X-ray sources in nearby galaxies regularly. Not only can we detect long-term X-ray variability, we can also find spectral variation as well as possible orbital period. I will review some of the important {\it Chandra} and {\it Swift} monitoring observations of nearby galaxies in the last decade. I will also discuss the possibility to use MAXI for X-ray monitoring observations of nearby galaxies.
}

\kword{black hole physics --- X-rays: binaries --- X-rays: galaxies}

\maketitle
\thispagestyle{empty}

\section{Introduction}

Over the last 20 years, all-sky monitoring observations of X-ray binaries in the Milky Way have resulted many important discoveries. For instance, detection of an outburst of an X-ray transient allows us to discover new black hole candidate, and subsequent localization at other wavelengths enables a detailed study about the nature of the system. By performing dynamical mass measurements via optical spectroscopy during the X-ray quiescent state after the intense outburst, we can obtain the mass function of the compact object to constrain its physical nature. By following the X-ray evolution of X-ray transients, one can study different intensity/spectral states and investigate the relationship with the nature of the compact object. In addition, some persistent X-ray binaries shows X-ray variability on timescales from milliseconds to years. While high timing resolution observations can probe variability from milliseconds to hours, orbital periods and super-orbital periods on timescales of days to years can only be studied with regular monitoring observations.

All these observations have proven to be very successful, but the sample is mainly limited to the Milky Way, which suffers difficulty in determining the distance to the source (hence uncertain luminosity measurements) and high extinction in some regions. With the improvement of the spatial resolution and sensitivity, monitoring X-ray sources in nearby galaxies has become feasible now. One intriguing discovery from X-ray observations of nearby galaxies is ultraluminous X-ray sources (ULXs), luminous ($L_X > 10^{39}$\lum) nonnuclear X-ray point-like sources in galaxies with apparent X-ray luminosities above the Eddington limit for a typical stellar-mass ($\sim 10 M_\odot$) black hole. The majority of ULXs are believed to be accreting objects in binary systems due to their strong X-ray flux variability observed on timescales of hours to years. Assuming an isotropic X-ray emission, ULX is the best candidate of intermediate-mass black hole (IMBH) with a mass of $\sim 10^2-10^4 M_\odot$ (Makishima et al. 2000; Miller \& Colbert 2004). While ULX may represent a missing link between stellar-mass black hole and supermassive black hole in galactic center, its formation and evolution is not well understood. 

The X-ray spectral properties may provide some hints about the connection between ULXs and Galactic black hole X-ray binaries. While Galactic black hole binaries have a good correlation between the spectral shape and luminosity, the behaviors of ULXs are more complex (see Gladstone et al. 2011 for a review). ULXs seem to require more complicated accretion geometry invoking outflows, corona, massive donors, and super-Eddington accretion flows (e.g. Stobbart et al. 2006; Poutanen et al. 2007; Patruno \& Zampieri 2008; Gladstone et al. 2009). It is possible that ULX is a distinct class of systems comparing to Galactic black hole binaries. Some very luminous ULXs may indeed have an IMBH accretor. Alternatively, ULX could be a typical stellar-mass black hole with geometrically or relativistically beamed emission (King et al. 2001; K\"ording et al. 2002) so that the X-ray luminosity does not exceed the Eddington limit.

Unlike Galactic X-ray binaries, X-ray sources (including ULXs) in nearby galaxies (apart from the Magellanic Clouds) cannot be detected with typical all-sky monitor instruments because of their X-ray faintness (see \S\,4 for an example). However, regular pointed X-ray observations can be employed as a monitoring instruments. In the past 10 years, {\it RXTE}, \chandra\ and \swift\ have been used for this purpose.
I here review some of the monitoring X-ray observations of nearby galaxies  using \chandra\ and \swift.

\section{Monitoring Observations of M31}
At a distance of 780 kpc, M31 provides us with a prime opportunity to study the global properties of a galaxy that is similar to our Galaxy. Hence, it is not surprising that almost all the X-ray instruments have observed M31. Even though many early X-ray instruments do not have imaging capability, M31 may be detected with an X-ray instrument as early as 1971 and a secure detection was made in 1973 (Giacconi et al. 1972; Bowyer et al. 1974). Table 1 lists all the X-ray observations of M31 before the \chandra\ and \xmm\ era. High resolution and sensitivity \chandra\ and \xmm\ observations have revolutionized the study of X-ray populations of nearby galaxies. In particular, \chandra\ has a long-term ongoing monitoring program to look for X-ray transients in M31 for which X-ray observations have been performed regularly (from every month to every few months; see Kong et al. 2002 for a description). \xmm\ and \swift\ also has a similar program to monitor nova in M31 (Henze et al. 2010). At the time of writing, there are already 1.2 Msec \chandra\ observations with 125 pointings, and 35 \xmm\ observations with a total of 835 ksec for the core of M31. The X-Ray Telescope (XRT) onboard \swift\ has also observed M31 for 78 times with over 250 ksec exposure.

These surveys have discovered over 60 X-ray transients and many X-ray variables in M31 (e.g., Kong et al. 2002; Di\,Stefano et al. 2004; Williams et al. 2004,2006). On average, there is one transient per month in M31 (Williams et al. 2006), higher than that in the Milky Way ($\sim2.6$ yr$^{-1}$; Chen et al. 1997). The lightcurves show similar profile and peak luminosities (Kong et al. 2002; Williams et al. 2006) comparing to those seen in Galactic X-ray transients. Moreover, by combining with the {\it Hubble Space Telescope}, we have discovered optical counterparts of a few X-ray transients in M31 (e.g., Williams et al. 2005). Using \chandra, \xmm\ and \swift, the only known classical novae in a M31 globular cluster was identified (Henze et al. 2009). The high spatial resolution of \chandra\ also allows us to discover an outburst of the central supermassive black hole in M31 (Li et al. 2011). 

Regular monitoring observations of M31 have become a routine operation for \chandra\ and indeed it has proved to be very successful in studying the X-ray populations in M31. Combining with the rapid response of \swift\ for TOO follow-up and all-sky ground-based optical surveys, we expect to discover more intriguing events in the near future. 

\begin{table}[t]
\caption{Historical X-ray Observations of M31 before {\it Chandra}}
\begin{center}
\begin{tabular}{p{2.6cm}p{5.3cm}} \hline\hline\\[-6pt]
Instrument       & Reference  \\   \hline
{\it Uhuru} & Giacconi et al. 1972$^*$; Forman et al. 1978 \\
Rocket-borne proportional counter & Bowyer et al. 1974  \\
{\it Ariel 5} & Cooke et al. 1978$^*$; McHardy et al. 1981 \\
{\it HEAO 1} & Wood et al. 1984 \\
{\it Einstein} & van Speybroeck et al. 1979 \\
{\it EXOSAT} & McKechnie et al. 1985 \\
{\it Ginga} & Makishima et al. 1989 \\
{\it ROSAT} & Primini et al. 1993 (HRI); Supper et al. 1997,2001 (PSPC)\\
{\it ASCA} & Takahashi et al. 2001 \\
{\it BeppoSAX} & Trinchieri et al. 1999 \\
{\it RXTE} & Revnivtsev et al. 2004 \\   \hline 
\end{tabular}
\end{center}
NOTE: * Probable detection
\end{table}

\section{{\it Swift} Monitoring Observations of Nearby Galaxies}

As described in \S\,1, the only way to monitor X-ray sources in nearby galaxies is through regular pointed observations with sensitive instruments. \chandra\ has been very successful in observing M31 regularly (see \S\,2), while {\it RXTE}/PCA discovered a 62-day period of the ULX, M82 X-1, presumably due to orbital modulation (Kaaret et al. 2006; Kaaret \& Feng 2007).
Although \swift/XRT is not an all-sky monitoring instrument, it has been used to perform snapshot observations on a regular basis for a few nearby galaxies hosting ULXs. Here we summarize the results of a few interesting ULXs as part of this program.

\subsection{NGC\,5408 X-1}

\ngc\ is one of the most luminous ULXs with an X-ray luminosity of $\sim 2\times10^{40}$\lum\ and it is among the few ULXs for which low frequency quasi-periodic oscillations have been found (Strohmayer et al. 2007; Strohmayer \& Mushotzky 2009). The source is also found to have a radio nebula (Kaaret et al. 2003) and a photoionized nebula (Kaaret \& Corbel 2009). These provide strong evidence that the compact object of NGC\,5408 X--1 is an IMBH. In spite of this, Middleton et al. (2011) challenged the IMBH suggestion by re-analyzing the \xmm\ data, and suggested a mass limit of $\laeq 100 M_\odot$.  This source is also one of the first ULXs observed regularly with \swift. Using the XRT data spanning $\sim 500$ days, Strohmayer (2009) discovered a 115 day periodicity and suggested that it is due to orbital modulation. If this interpretation is correct, it would support that NGC\,5408 X-1 contains a $\sim1000 M_\odot$ black hole with a 3-$5 M_\odot$ companion. However, Foster et al. (2010) argued that the modulation is due to super-orbital period (see Charles et al. 2008) implying that the system has a shorter orbital period and likely contains a stellar-mass black hole. Continued monitoring observations will be able to test the stability of the modulation and confirm if the period is truly orbital in origin.

Here we used more updated data to search for periodicity. Using \swift/XRT data covering over 1000 days, we plot the long-term lightcurve in Figure 1. Modulation on a timescale of $\sim100$ days is evident. We performed a Lomb-Scargle periodogram analysis and the resulting power spectrum is shown in Figure 1. The highest peak is at 118 days, consistent with the $115\pm4$ day periodicity found by Strohmayer (2009). Moreover, there are several dip-like features with a separation of $\sim200$ days which is shown as the second highest peak in the periodogram. While the 200-day periodicity is not statistically significant, the physical nature of these dips is unclear; further monitoring observations are required to confirm the periodicity.

%
%
\begin{figure}[t]
\centering
\psbox[xsize=9.1cm]{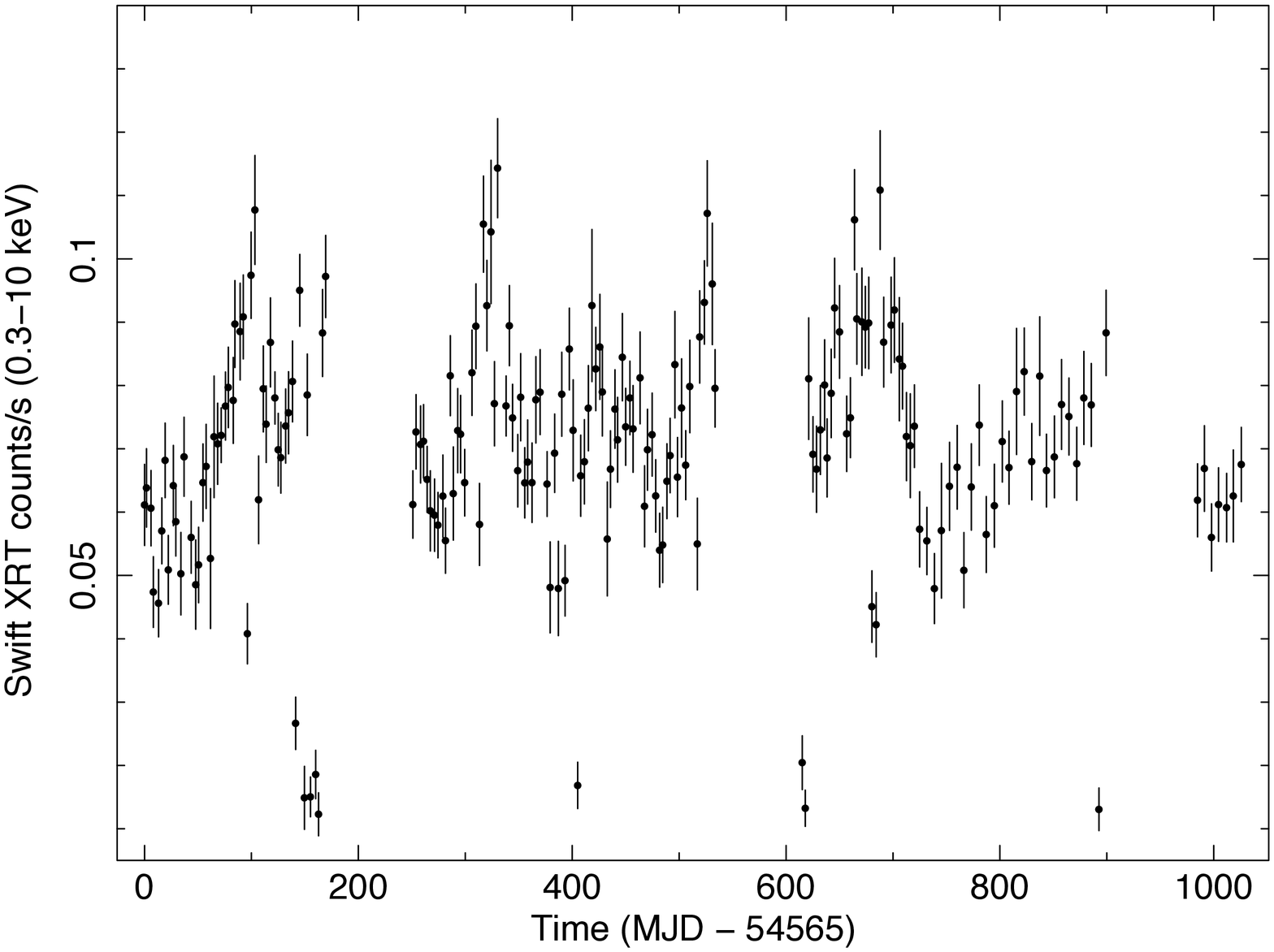}
\vspace{-1.5cm}
\psbox[xsize=9.1cm]{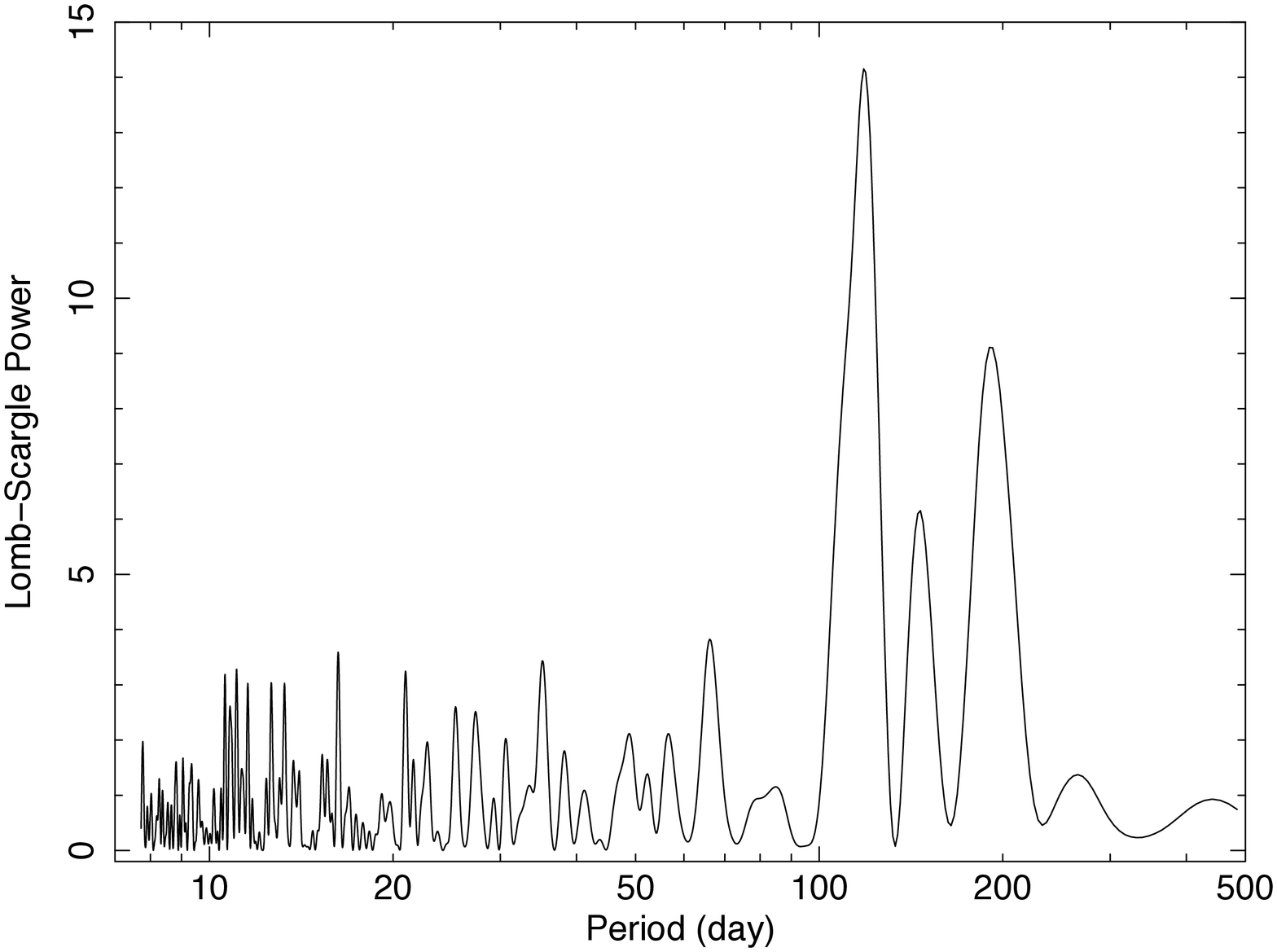}
\caption{Long-term lightcurve (upper panel) and Lomb-Scargle periodogram (lower panel) of NGC\,5408 X-1 using {\it Swift}/XRT data. The highest peak of the periodogram is at 118 days. The 99.9\% significance level has a power of 12.4.}
\end{figure}

\subsection{ESO\,243-49 HLX-1}

The ULX \eso\ HLX-1 is the most luminous ULX that has a peak X-ray luminosity of  $1.2\times10^{42}$\lum\ (Farrell et al. 2009). Assuming an isotropic radiation, the black hole mass is estimated to be $\sim 500 M_\odot$. By using the position provided by \chandra, Soria et al. (2010) identified a faint optical counterpart, and subsequent optical spectroscopy with the Very Large Telescope confirmed that the source is located in \eso\ with a redshift of $z=0.0223$ (Wiersema et al. 2010). The $10^{42}$\lum\ luminosity of the source is therefore without doubt, strongly supporting a $500 M_\odot$ IMBH accretor. 

\eso\ HLX-1 is thus an ideal target for monitoring observations. \swift\ has begun a monitoring program for the source since late-2008. The long-term lightcurve of the source is shown in Figure 2. It is clear that the source exhibits substantial variability. In particular, there are two outbursts with a profile similar to typical Galactic X-ray transients. We show in Figure 2 the lightcurves derived from two energy bands (0.3--1.5 keV and 1.5--10 keV). It is evident that the energy spectra were changing during an outburst. The source seems to be softer during the peak of the outburst. It is similar to Galactic X-ray transients transiting from the low/hard state to the high/soft state. The spectral variation of the first outburst was already noticed by Godet et al. (2009). After the source returned to the low state, the second outburst occurred about 200 days later. Like the first one, the lightcurve has a fast-rise-exponential-decay profile and the spectrum softened during the peak of the outburst.

Based on the limited data, the two outbursts seem to be separated by about 400 days. By performing a Lomb-Scargle periodogram analysis, the strongest peak is at 373 days. The 99.9\% significance level has a power of 11.2, indicating that the period is statistically significant. There are a few data points taken in 2008 October/November and they hint that the 373-day periodicity may be real. If it is the case, the next outburst will be around 2011 August, which can be tested by future observations.

\begin{figure}[t]
\centering
\psbox[xsize=9.1cm]{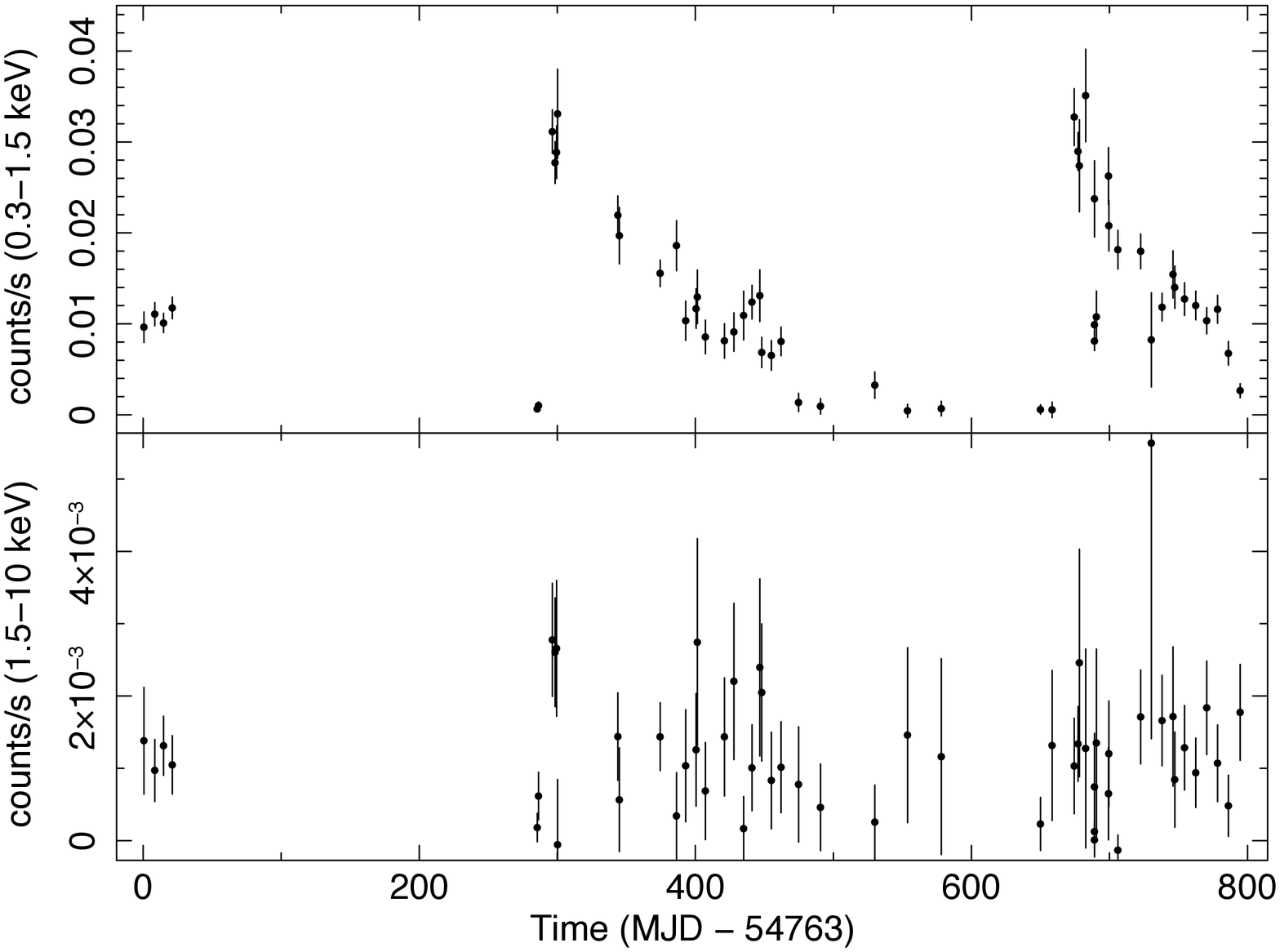}
\vspace{-1.5cm}
\psbox[xsize=9.1cm]{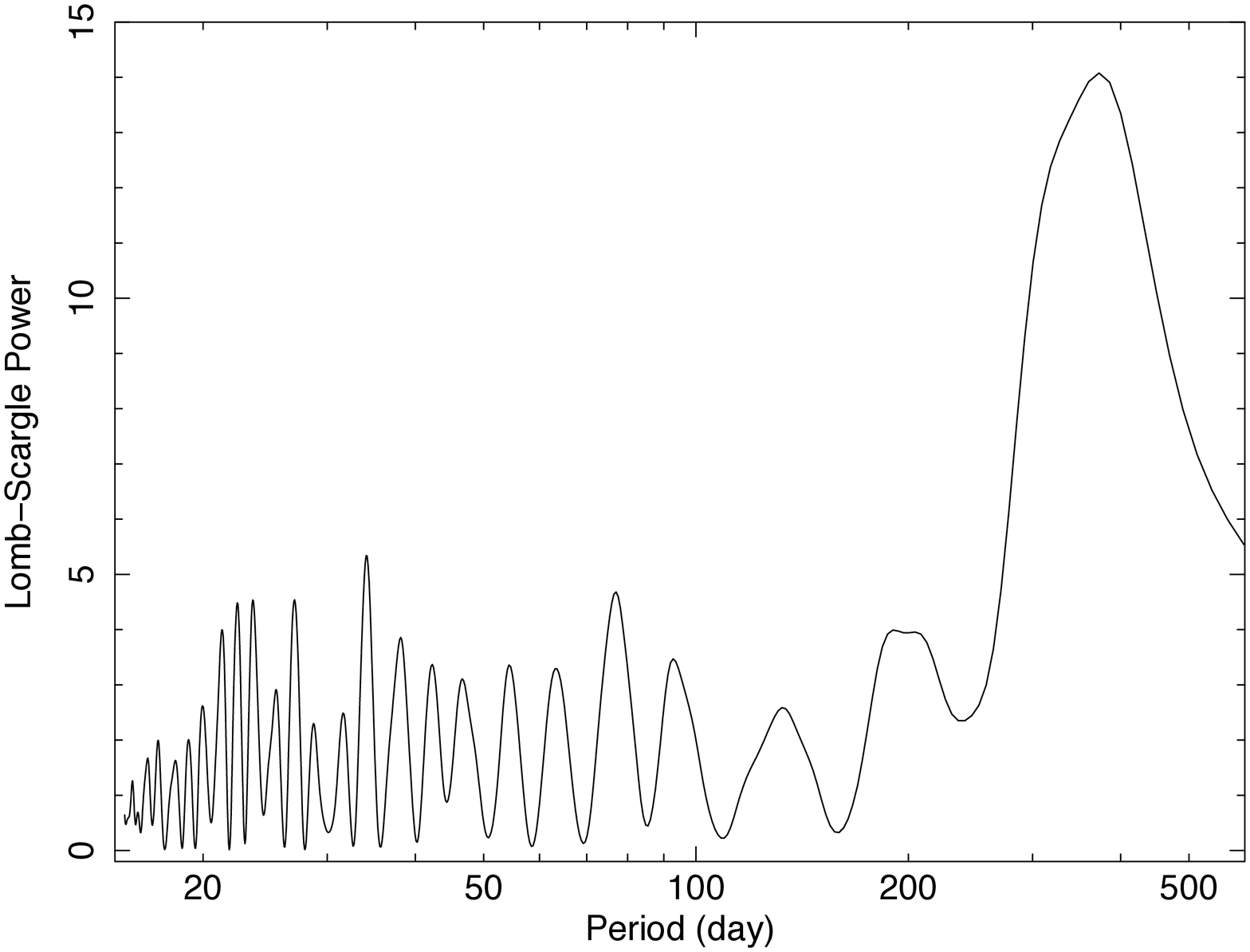}
\caption{Long-term lightcurve (upper panel) and Lomb-Scargle periodogram (lower panel) of \eso\ HLX-1 using {\it Swift}/XRT data. The lightcurve is divided in to two energy bands (0.3--1.5 keV and 1.5--10 keV) to show the spectral evolution. It is clear that the outbursts are only seen in the soft X-ray band indicating that the spectra softened during the outbursts. The highest peak of the periodogram is at 373 days. The 99.9\% significance level has a power of 11.2.}
\end{figure}

\subsection{Holmberg IX X-1}
\ho\ is a well-known ULX with an X-ray luminosity of $\sim10^{40}$\lum\ located near the dwarf companion of M81, Holmberg IX. The source was first discovered by the {\it Einstein Observatory} (Fabbiano 1988) and has been observed by all major X-ray observatories throughout the last 20 years (La Parola et al. 2001). Apart from the X-ray flux variability, \ho\ is also one of the first ULXs shown to have a cool ($\sim 0.1-0.2$ keV) accretion disk, leading to a suggestion of an IMBH accretor (Miller et al. 2004). Recently, Kong et al. (2010) used \swift\ to monitor the X-ray evolution of \ho\ and utilized the co-added spectra taken at different luminosity states to study the spectral behavior of the source. The best spectral fits are provided by a dual thermal model with a cool blackbody and a warm disk blackbody. This suggests that \ho\ may be a $10 M_\odot$ black hole accreting at seven times above the Eddington limit or a $100 M_\odot$ maximally rotating black hole accreting at the Eddington limit, and we are observing both the inner regions of the accretion disk and outflows from the compact object. 

Here we provide a more updated lightcurve of \ho. Figure 3 shows the long-term lightcurve and the Lomb-Scargle periodogram. Comparing to Kong et al. (2010), \ho\ has gone through another high/low intensity transition. The two period candidates ($\sim 20$ and 60 days; Kaaret \& Feng 2009 and Kong et al. 2010) in previous analyses are not statistically significant. Instead, a period of 208 days is above the 99.9\% significance level. Since the lightcurve is dominated by the two high/low intensity transitions, we do not interpret the periodicity is real at this moment. Further monitoring observations are required to confirm the modulation.

\subsection{M81 X-6}
M81 X-6 is the brightest non-nuclear X-ray source in M81. The average X-ray luminosity of the source is $\sim2\times10^{39}$\lum. M81 X-6 is identified with a single blue star by using the {\it Hubble Space Telescope} (Liu et al. 2002). The optical properties are consistent with an early-type main-sequence star. Based on X-ray spectroscopy, the black hole mass is estimated to be $18 M_\odot$ (Swartz et al. 2003). Liu et al. (2002) suggested that M81 X-6 is a high-mass X-ray binary with a $18 M_\odot$ black hole and a $23 M_\odot$ O8 V companion, and predicted that the orbital period is about 1.8 days. Although the cadence of the current \swift\ monitoring program for ULXs is usually from a few days to a week and such a short orbital period is difficult to be detected, we can test if a longer period exists particularly when the system geometry is not well constrained at the moment. 

In Figure 4, we show the \swift\ long-term lightcurve of M81 X-6 and its Lomb-Scargle periodogram. The source has varied by a factor of $\sim4$ on a timescale of days. Period analysis indicates that there is no significant modulation between 6 and 300 days although a period of 175 days is near the significance level. Continuous monitoring observations will be able to confirm if the period is real.

\begin{figure}[t]
\centering
\psbox[xsize=9.1cm]{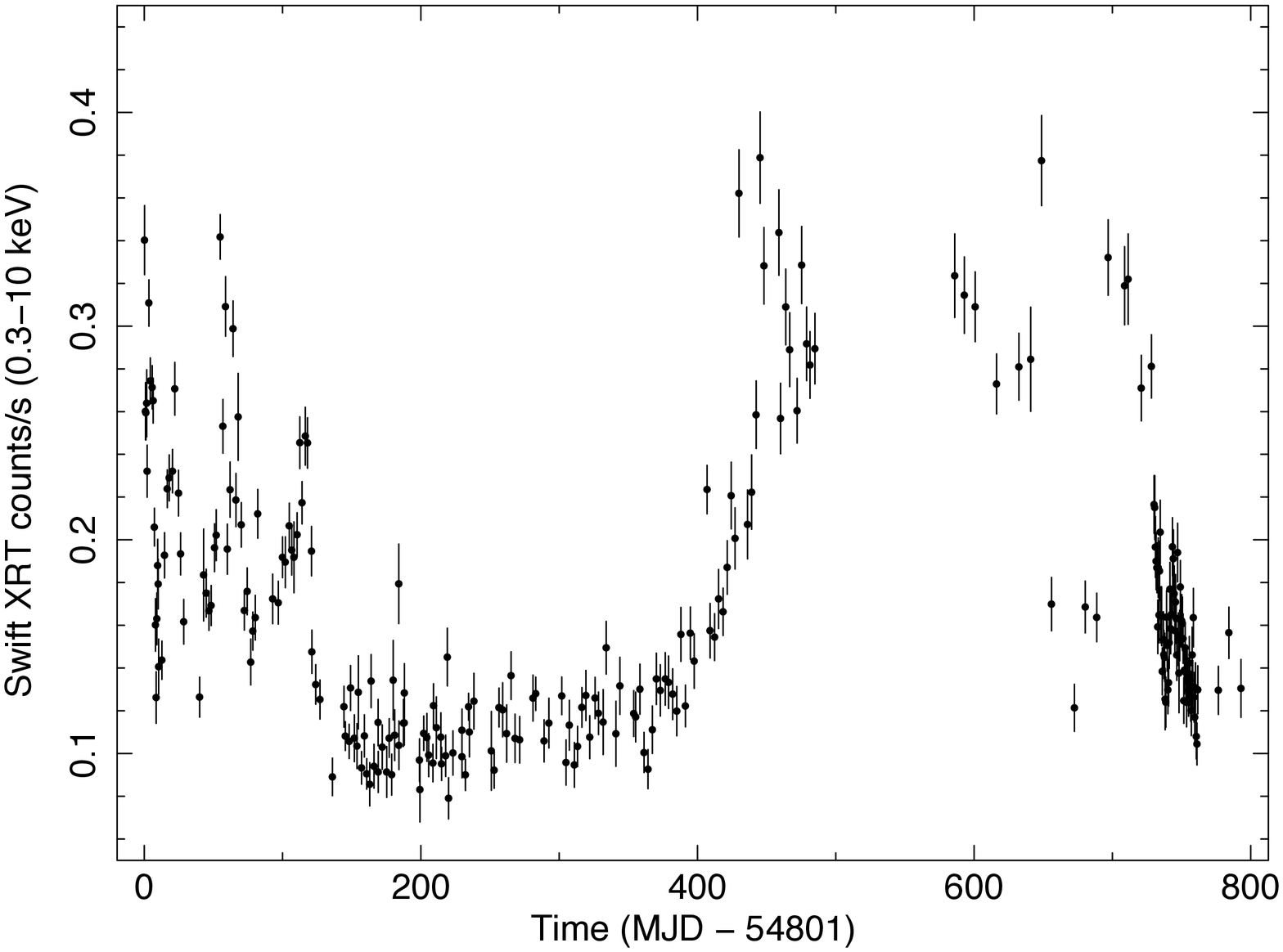}
\vspace{-1.5cm}
\psbox[xsize=9.1cm]{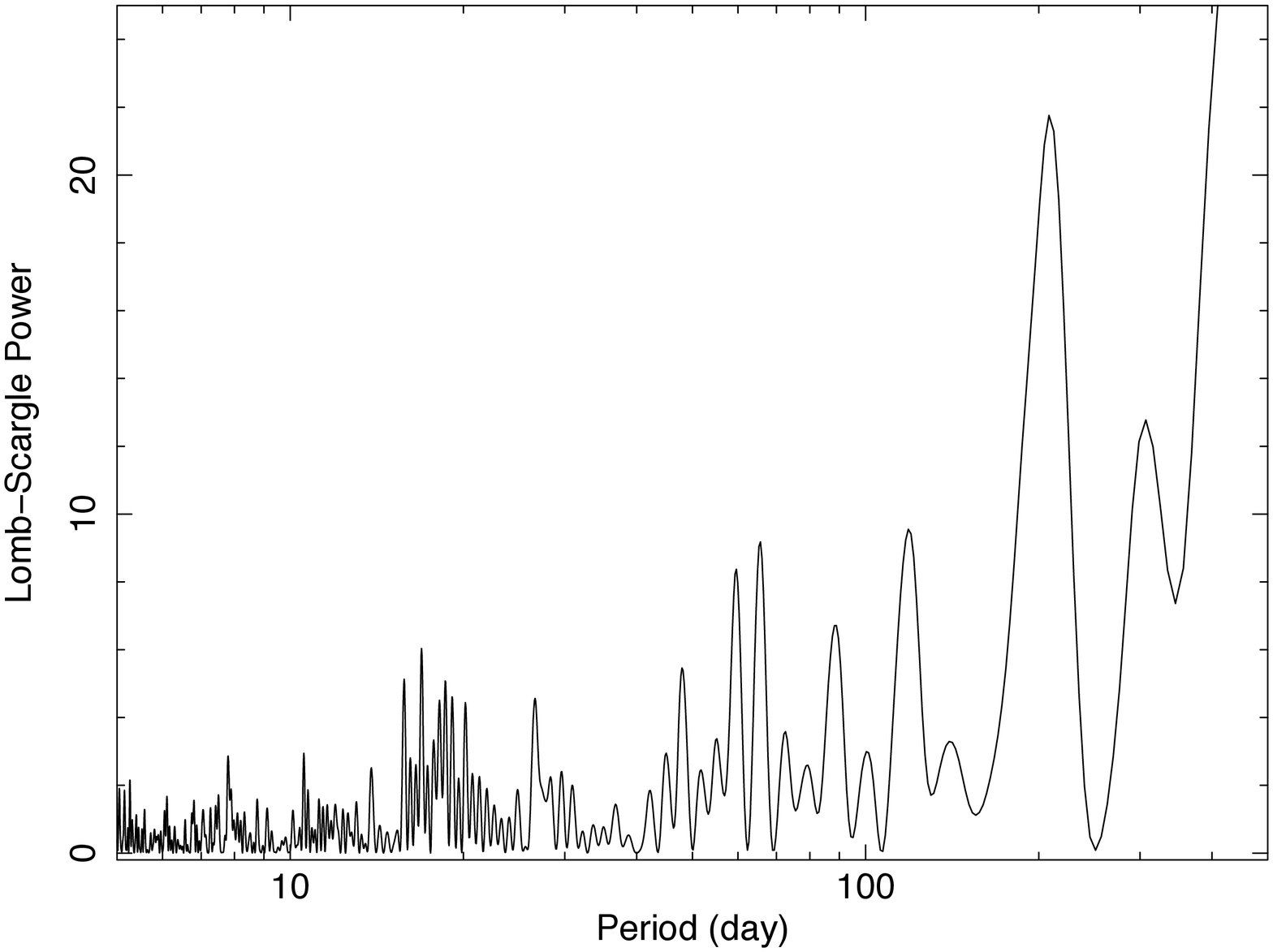}
\caption{Long-term lightcurve (upper panel) and Lomb-Scargle periodogram (lower panel) of \ho\ using {\it Swift}/XRT data. The highest peak of the periodogram is at 208 days. The 99.9\% significance level has a power of 12.7.}
\end{figure}

\begin{figure}[t]
\centering
\psbox[xsize=9.1cm]{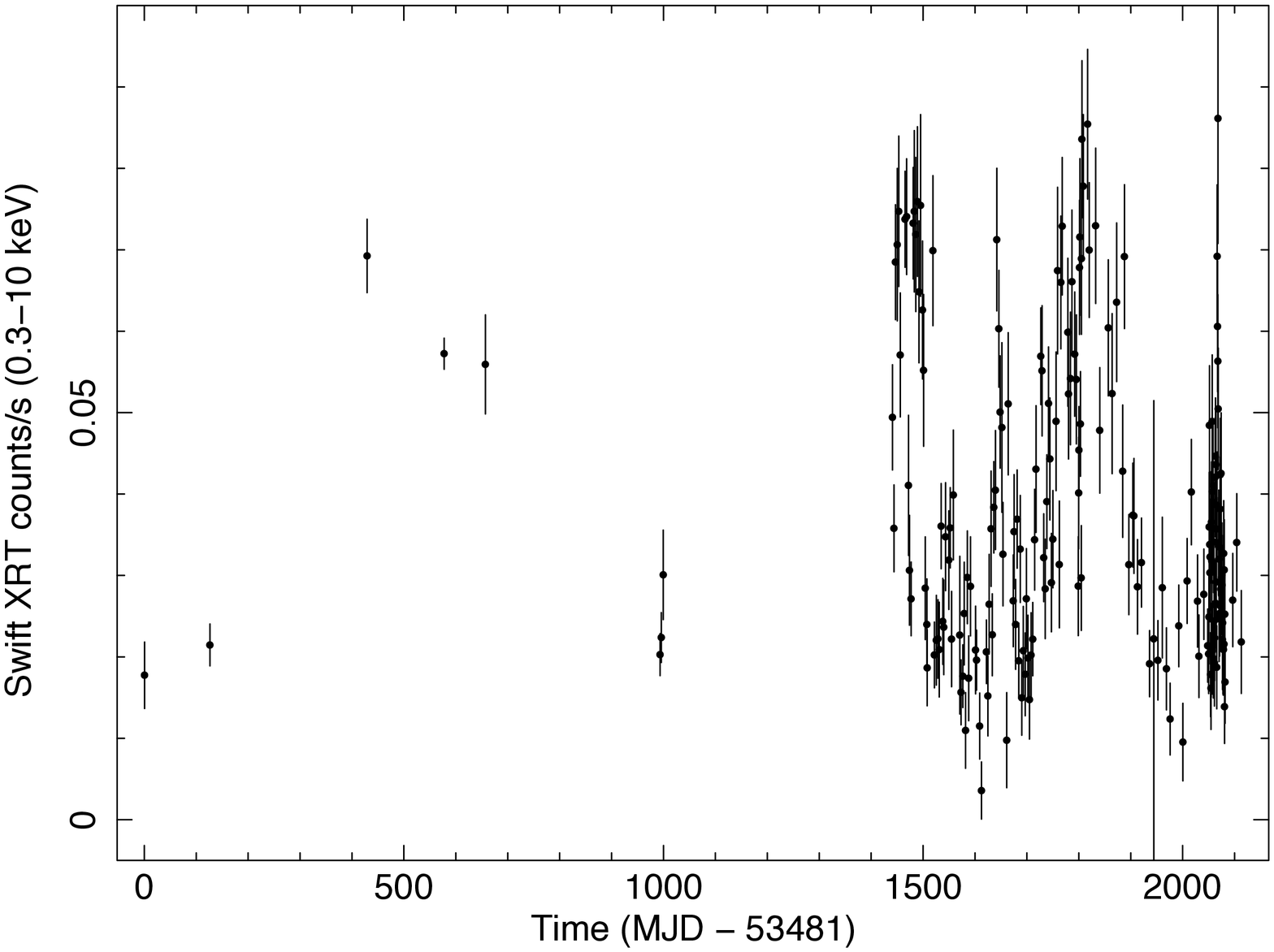}
\vspace{-1.5cm}
\psbox[xsize=9.1cm]{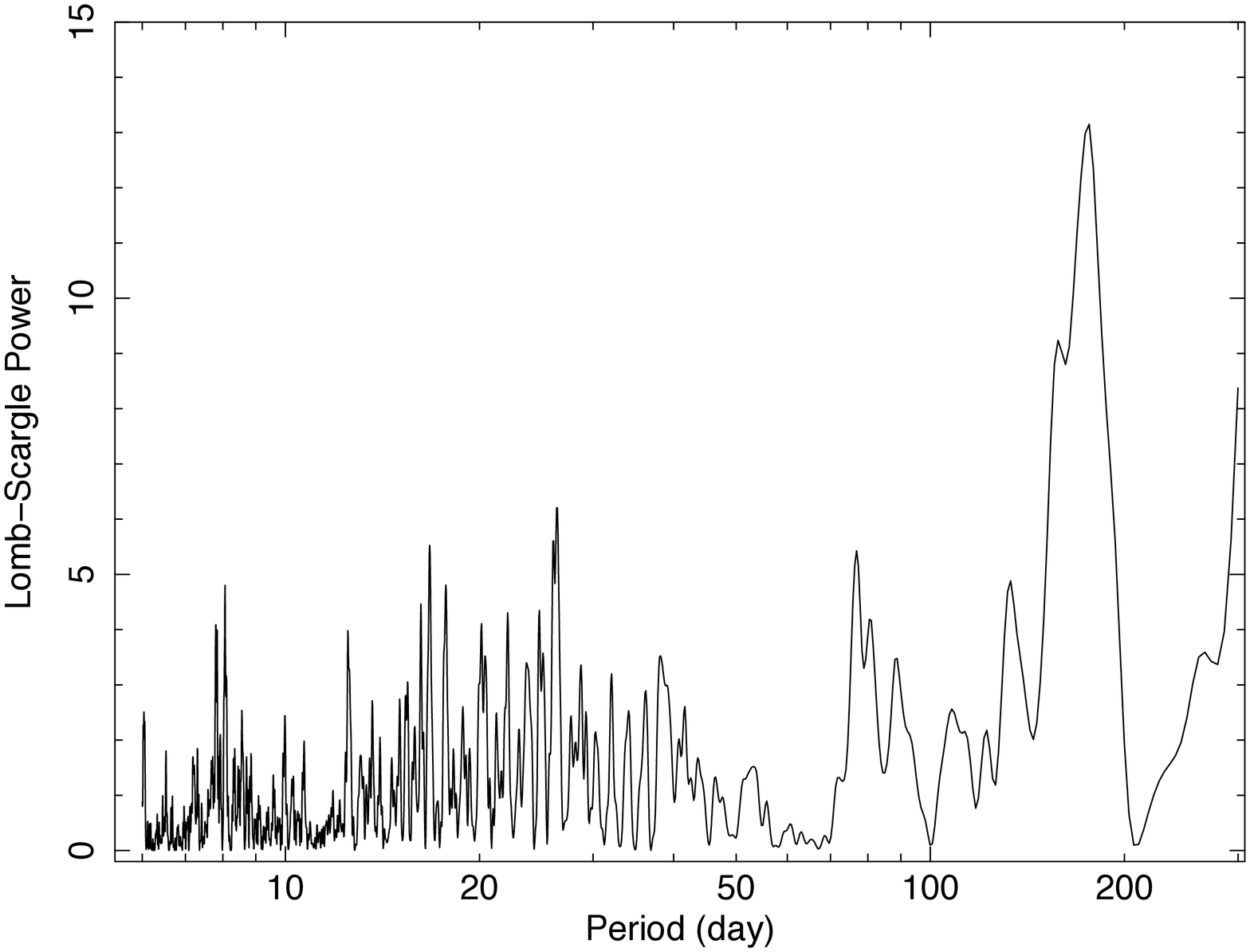}
\caption{Long-term lightcurve (upper panel) and Lomb-Scargle periodogram (lower panel) of M81 X-6 using {\it Swift}/XRT data. The highest peak of the periodogram is at 175 days. The 99.9\% significance level has a power of 13.}
\end{figure}

\section{Prospects of MAXI in Monitoring Variable X-ray Sources of Nearby Galaxies}
Monitoring X-ray sources located in nearby galaxies (except for the Magellanic Clouds) using all-sky instruments like MAXI is challenging because of their sensitivity. Based on the current performance of MAXI/GSC, the daily $5\sigma$ sensitivity is 15 mCrab which is three times worse than that of the pre-flight simulation (Sugizaki et al. 2011). Even with a monthly integration, the sensitivity is only about 2.7 mCrab. The nearest spiral galaxy, M31, is about 2.2 mCrab in the 2--20 keV band (see e.g., Makishima et al. 1989) which is slightly below the detection limit. In principle, M31 should be detectable by integrating 2 months of data. However, with a 2-month cadence, it is unlikely that we can detect any transients in M31 if they are similar to those in the Milky Way. Nevertheless, M31 is indeed detected with MAXI/GSC at $16\sigma$ level by integrating 7 months of data and the averaged flux is $3.04\pm0.19$ mCrab (Hiroi et al. 2011) based on preliminary analysis.

The next possible target for MAXI is M82 where an ULX (M82 X-1) with a peak X-ray luminosity of $10^{41}$\lum\ is located. At a distance of 3.6 Mpc, the corresponding flux will be slightly smaller than M31. However, M82 X-1 is known as a variable with a factor of $> 20$ variability (Chiang \& Kong 2011). Therefore we expect to detect M82 with at least 3 months of data. One advantage of M82 however is that the galaxy's flux is dominated by M82 X-1; we can therefore make use of MAXI to monitor the variability of M82 X-1 on a timescale of a few months.

If the sensitivity of MAXI can be improved by a factor of $> 2$ in the future, a more meaningful monitoring survey of M31 and M82 can be performed. This also suggests that future soft X-ray all-sky monitoring instruments will be very useful to study nearby galaxies if the sensitivity can reach $\laeq 1$ mCrab per day.

\vspace{1pc}
\noindent I would like to thank K. Hiroi for useful discussion and providing the MAXI measurement of M31. This work made use of data supplied by the UK Swift Science Data Centre at the University of Leicester. This project is supported by the National Science Council of the Republic of China (Taiwan) through grants NSC96-2112-M-007-037-MY3 and 
NSC99-2112-M-007-004-MY3. I would like to acknowledge support from a Kenda Foundation Golden Jade Fellowship of Taiwan.

\section*{References}

\re
Bowyer, S. et al. 1974 ApJ, 190, 285

\re
Charles, P. et al. 2008 NewAR, 51, 768

\re
Chen, W. et al. 1997 ApJ, 491, 312

\re
Chiang, Y.-K. \& Kong, A.K.H. 2011 MNRAS, in press, arXiv1101.0414

\re
Cooke, B.A. et al. 1978 MNRAS, 182, 489

\re
Di Stefano, R. et al. 2004 ApJ, 610, 247

\re
Fabbiano, G. 1988 ApJ, 325, 544

\re
Farrell, S.A. et al. 2009 Nature, 460, 73

\re
Forman, W. et al. 1978 ApJS, 38, 357

\re
Foster, D.L. et al. 2010 ApJ, 725, 2480

\re
Giacconi, R. et al. 1972 ApJ, 178, 281

\re
Gladstone, J.C. et al. 2009 MNRAS, 397, 1836

\re
Gladstone, J.C. et al. 2011 Astronomische Nachrichten, arXiv:1101.4859

\re
Godet, O. et al. 2009 ApJ, 705, L109

\re
Henze, M. et al. 2009 A\&A, 500, 769

\re
Henze, M. et al. 2010 A\&A, 523, 89

\re
Hiroi, K. et al. 2011 in this proceedings

\re
Kaaret, P. et al. 2003 Science, 299, 365

\re
Kaaret, P. et al. 2006 Science, 311, 491

\re
Kaaret, P. \& Feng, H. 2007 ApJ, 669, 106

\re
Kaaret, P. \& Feng, H. 2009 ApJ, 702, 1679

\re
Kaaret, P. \& Corbet, S. 2009 ApJ, 697, 950

\re
King, A.R. et al. 2001 ApJ, 552, L109

\re
Kong, A.K.H. et al. 2002 ApJ, 577, 738

\re
Kong, A.K.H. et al. 2010 ApJ, 722, 1816

\re
K\"ording, E. et al. 2002 A\&A, 382, L13

\re
La Parola, V. et al. 2001 ApJ, 556, 47

\re
Li, Z. et al. 2011 ApJ, 728 10

\re
Liu, J. et al. 2002 ApJ, 580, L31

\re
Makishima, K. et al. 1989 PASJ, 41, 697

\re
Makishima, K. et al. 2000 ApJ, 535, 632

\re
McHardy, I.M. et al. 1981 MNRAS, 197, 893

\re
McKechnie, S.P. et al. 1985 in X-ray astronomy '84, p.373

\re
Middleton, M.J. et al. 2011 MNRAS, 411, 644

\re
Miller, J. M. et al. 2004 ApJ, 614, L117

\re
Miller, M.C., \& Colbert, E.J.M. 2004 Int. J. Mod. Phys. D, 13, 1

\re
Patruno, A., \& Zampieri, L. 2008 MNRAS, 386, 543

\re
Poutanen, J. et al. 2007 MNRAS, 377, 1187

\re
Primini, F.A. et al. 1993 ApJ, 410, 615

\re
Revnivtsev, M. et al. 2004 A\&A, 418, 927

\re
Soria, R. et al. 2010 MNRAS, 405, 870

\re
Stobbart, A.-M. et al. 2006 MNRAS, 368, 397

\re
Strohmayer, T.E., Mushotzky, R.F. 2009 ApJ, 703, 1366

\re
Strohmayer, T.E 2009 ApJ, 706, L210

\re
Sugizaki, M. et al. 2011 PASJ, in press, arXiv:1102.0891

\re
Supper, R. et al. 1997 A\&A, 317, 328

\re
Supper, R. et al. 2001 A\&A, 373, 63

\re
Swartz, D.A. et al. 2003 ApJS, 144, 213

\re
Takahashi, H. et al. 2001 PASJ, 53, 875

\re
Trinchieri, G. et al. 1999 A\&A, 348, 43


\re
van Speybroeck, L. 1979 ApJ, 234, 45

\re
Wiersema, K. et al. 2010 ApJ, 721, L102

\re
Williams, B.F. et al. 2004 ApJ, 609, 735

\re
Williams, B.F. et al. 2005 ApJ, 620, 723

\re
Williams, B.F. et al. 2006 ApJ, 643, 356

\re
Wood, K.S. 1984 ApJS, 56, 507

\label{last}

\end{document}